\documentclass[preprint,3p]{elsarticle}
\usepackage{amssymb}
\usepackage{amsthm}
\usepackage[mathlines]{lineno}
\usepackage{graphicx}
\usepackage{units}
\usepackage{url}
\usepackage{amsmath}
\usepackage{amsfonts}
\usepackage{bm}
\usepackage{textcomp}
\usepackage{subfigure}
\usepackage{multicol}
\usepackage{verbatim}
\usepackage{rotating}
\usepackage[colorlinks,linkcolor=red,citecolor=red]{hyperref}
\usepackage{float}
\usepackage{epsfig}
\usepackage{dcolumn}
\usepackage{bm}
\usepackage{color}
\usepackage{pstricks}
\usepackage{pst-node}
\usepackage{times}
\usepackage{indentfirst}
\usepackage[english]{babel}
\addto{\captionsenglish}{%

}
\journal{Physics Letters B}


\begin{document}
\begin{frontmatter}
\title{{\bf
\boldmath Measurements of the branching fractions for $D^+\to
K^0_SK^0_SK^+$, $K^0_SK^0_S\pi^+$ and $D^0\to K^0_SK^0_S$,
$K^0_SK^0_SK^0_S$}}

\author{
\begin{small}
\begin{center}
M.~Ablikim$^{1}$, M.~N.~Achasov$^{9,e}$, S. ~Ahmed$^{14}$,
X.~C.~Ai$^{1}$, O.~Albayrak$^{5}$, M.~Albrecht$^{4}$,
D.~J.~Ambrose$^{44}$, A.~Amoroso$^{49A,49C}$, F.~F.~An$^{1}$,
Q.~An$^{46,a}$
% ...
\\
\vspace{0.2cm}
(BESIII Collaboration)\\
\vspace{0.2cm} {\it
$^{1}$ Institute of High Energy Physics, Beijing 100049, People's Republic of China\\
% ...
$^{4}$ Bochum Ruhr-University, D-44780 Bochum, Germany\\
$^{5}$ Carnegie Mellon University, Pittsburgh, Pennsylvania 15213, USA\\
% ...
$^{9}$ G.I. Budker Institute of Nuclear Physics SB RAS (BINP), Novosibirsk 630090, Russia\\
% ...
$^{14}$ Helmholtz Institute Mainz, Johann-Joachim-Becher-Weg 45, D-55099 Mainz, Germany\\
% ...
$^{44}$ University of Rochester, Rochester, New York 14627, USA\\
% ...
$^{46}$ University of Science and Technology of China, Hefei 230026, People's Republic of China\\
% ...
$^{49}$ (A)University of Turin, I-10125, Turin, Italy; (B)University of Eastern Piedmont, I-15121, Alessandria, Italy; (C)INFN, I-10125, Turin, Italy\\
% ...
$^{a}$ Also at State Key Laboratory of Particle Detection and Electronics, Beijing 100049, Hefei 230026, People's Republic of China\\
% ...
$^{e}$ Also at the Novosibirsk State University, Novosibirsk, 630090, Russia\\
% ...
}\end{center}
\vspace{0.4cm}
\end{small}
}

\begin{abstract}
By analyzing $2.93\ \rm fb^{-1}$ of data taken at the $\psi(3770)$
resonance peak with the BESIII detector, we measure the branching
fractions for the hadronic decays $D^+\to K^0_SK^0_S K^+$, $D^+\to K^0_SK^0_S \pi^+$,
$D^0\to K^0_S K^0_S$ and $D^0\to K^0_SK^0_SK^0_S$.
They are determined to be ${\mathcal B}(D^+\to K^0_SK^0_SK^+)=(2.54 \pm 0.05_{\rm
stat.} \pm 0.12_{\rm sys.})\times 10^{-3}$, ${\mathcal B}(D^+\to
K^0_SK^0_S\pi^+)=(2.70 \pm 0.05_{\rm stat.} \pm 0.12_{\rm
sys.})\times 10^{-3}$, ${\mathcal B}(D^0\to K^0_SK^0_S)=(1.67 \pm
0.11_{\rm stat.} \pm 0.11_{\rm sys.})\times 10^{-4}$ and ${\mathcal
B}(D^0\to K^0_SK^0_SK^0_S)=(7.21 \pm 0.33_{\rm stat.} \pm 0.44_{\rm
sys.})\times 10^{-4}$, where the second one is measured for the first time
and the others are measured with significantly improved precision over
the previous measurements.
\end{abstract}
\end{frontmatter}

\begin{multicols}{2}

\section{Introduction}
Hadronic decays of $D$ mesons open a window to probe for the physics mechanisms
in charmed meson decays, $e.g.$, CP violation,
$D^0\bar D^0$ mixing and SU(3) symmetry breaking effects.
Since the discovery of $D$ mesons in 1976, the hadronic decays of $D$ mesons
have been extensively investigated~\cite{pdg2014}. However, the
existing measurements of the $D$ hadronic decays containing 
at least two $K^0_S$ mesons in the final state are still very poor due to limited
statistics~\cite{pdg2014}.

In this Letter, we report the measurements of the branching fractions for
the hadronic decays $D^+\to K^0_SK^0_S \pi^+$, $D^0\to K^0_S K^0_S$,
$D^+\to K^0_SK^0_S K^+$ and $D^0\to K^0_SK^0_SK^0_S$.
Throughout this Letter,
charged conjugate modes are implied.
These decays have simpler event topologies and suffer less
from combinatorial backgrounds than other decay modes containing two
$K_S^0$ in the final state.
The comprehensive or improved measurements of three-body decays will benefit the understanding
of the interplay between the weak and strong interactions in multibody decays
where theoretical predictions are poorer than two-body decays.
The improved measurements of two-body decays can serve to better explore
the contributions of W-exchange diagrams and final-state interactions~\cite{plb193_331,prd45_4113,prd60_014014,prd64_034010},
as well as SU(3)-flavor symmetry breaking effects~\cite{Kwong,prd86_036012,Grossman,prd92_014004,prd92_014032}
in $D$ meson decays.
In addition, these measurements will also help to improve background estimations
in the precision measurements of $D$ and $B$ meson decays.

The data 
sample used for this analysis, which
has an integrated luminosity of $2.93$
fb$^{-1}$~\cite{lum}, was taken at the $\psi(3770)$ resonance peak
with the BESIII detector~\cite{bes3}. The $D^0\bar D^0$ and $D^+D^-$
pairs produced in $\psi(3770)$ decay provide cleaner $D^0$ and $D^+$ meson
samples than those used in previous studies at ARGUS~\cite{zpc46_9,zpc64_375},
CLEO~\cite{prd44_3383,prd44_4211} and FOCUS~\cite{plb607_59}. To optimize
the precision for these measurements, we use a single-tag method, in which
either a $D$ or $\bar D$ is reconstructed in an event.  We combine the yields measured 
with previously reported values of the cross sections for $e^+e^-\to D^0\bar
D^0$ and $D^+D^-$ at the $\psi(3770)$ resonance peak~\cite{crsdd-cleo}.

\section{BESIII detector and Monte Carlo simulation}
\label{sec:detector} The BESIII detector is a magnetic spectrometer 
that operates at the BEPCII collider.  It has a cylindrical geometry with a solid-angle 
coverage of 93\% of $4\pi$. It consists of several main components. A 43-layer
main drift chamber (MDC) surrounding the beam pipe performs precise
determinations of charged particle trajectories and measures the specific ionization 
($dE/dx$) for charged particle identification (PID). An array of time-of-flight 
counters (TOF) is located outside the MDC and provides additional PID
information. A CsI(Tl) electromagnetic calorimeter (EMC) surrounds
the TOF and is used to measure the energies of photons and
electrons. A solenoidal superconducting magnet outside the EMC
provides a 1 T magnetic field in the central tracking region of the
detector. The iron flux return of the magnet is instrumented with
1272 m$^2$ of resistive plate muon counters (MUC) arranged in
nine layers in the barrel and eight layers in the endcaps for identification of 
muons with momentum greater than 0.5 GeV/$c$. More
details about the BESIII detector are described in Ref.~\cite{bes3}.

A GEANT4-based \cite{geant4} Monte Carlo (MC) simulation software
package, which includes the geometric description and response of the 
detector, is used to determine the detection efficiency and
to estimate background for each decay mode.
An inclusive MC sample, which includes the $D^0\bar D^0$, $D^+D^-$ and non-$D\bar D$ decays of the
$\psi(3770)$, initial-state-radiation (ISR) production of the
$\psi(3686)$ and $J/\psi$, the $e^+e^-\to q\bar q$ ($q=u$, $d$, $s$) continuum
process, the Bhabha scattering events, the di-muon events and the
di-tau events, is produced at $\sqrt s=3.773$ GeV. The equivalent luminosity of the MC sample is ten times of data.
The $\psi(3770)$ decays are generated by the MC generator KKMC \cite{kkmc}, which
incorporates both ISR effects \cite{isr} and final-state-radiation (FSR) 
effects \cite{photons}. Known decay modes are generated using
EvtGen \cite{evtgen} with input branching fractions from the
Particle Data Group (PDG)~\cite{pdg2014}.  Unmeasured decays are 
generated using LundCharm \cite{lundcharm}.

\section{Data analysis}
\label{sec:evtsel}
All charged tracks used in this analysis are required to be within
a polar-angle ($\theta$) range of $|\rm{cos~\theta}|<0.93$.
The good charged tracks, except when used to reconstruct $K^0_{S}$
mesons, are required to originate within an interaction region defined by
$V_{xy}<$ 1.0 cm and $V_{z}<$ 10.0 cm,
where $V_{xy}$ and $V_{z}$ are the distances of closest approach
of the reconstructed track to the interaction point (IP) perpendicular to 
($xy$) and along ($z$) the beam direction.

The charged kaons and pions are
identified by the $dE/dx$ and TOF measurements.
The combined confidence levels
for pion and kaon hypotheses ($CL_{\pi}$ and $CL_{K}$) are calculated, respectively.
The charged track is identified as kaon (pion) if $CL_{K}>CL_{\pi}$ ($CL_{\pi}>CL_{K}$) is satisfied.

$K^0_{S}$ candidate mesons are reconstructed through the $\pi^{+}\pi^{-}$ decay mode.
Charged pions used in $K^0_{S}$ candidates mesons are required to
satisfy $V_{z}<$ 20.0 cm.
The two oppositely charged tracks are assumed to be a $\pi^+\pi^-$
pair without PID requirements. To reconstruct $K^0_S$,
the $\pi^+\pi^-$ combination is constrained to have a common vertex.
The candidate is accepted if it has an invariant mass $M_{\pi^+\pi^-}$ 
within $12$ MeV$/c^{2}$ of the $K^0_{S}$ nominal mass~\cite{pdg2014} and 
satisfies $L/\sigma_{L}>$ 2, where $L$ is the measured flight distance and
$\sigma_{L}$ is its uncertainty.

To identify $D$ candidates, we use two selection variables,
the energy difference $\Delta E \equiv E_D-E_{\rm beam}$ and
the beam-energy-constrained mass 
$M_{\rm BC} \equiv \sqrt{E^{2}_{\rm beam}/c^4-|\vec{p}_{D}|^{2}/c^2}$,
where $E_{\rm beam}$ is the beam energy and
$E_D$ and $\vec{p}_{D}$ are the energy and momentum
of the $D$ candidate in the $e^+e^-$ center-of-mass system.
For each signal decay mode, only the combination with
the minimum $|\Delta E|$ is kept in events where more than one candidate passes 
the selection requirements.
Mode-dependent $\Delta E$ cuts are determined separately for data and 
MC based on fits to the respective $\Delta E$ distributions.  These are set at 
$\pm 3\sigma$, where $\sigma$ is the $\Delta E$ resolution (Table~\ref{tab:deltaE}).

\begin{figure*}[htp]
  \centering
  \includegraphics[width=2.1in]{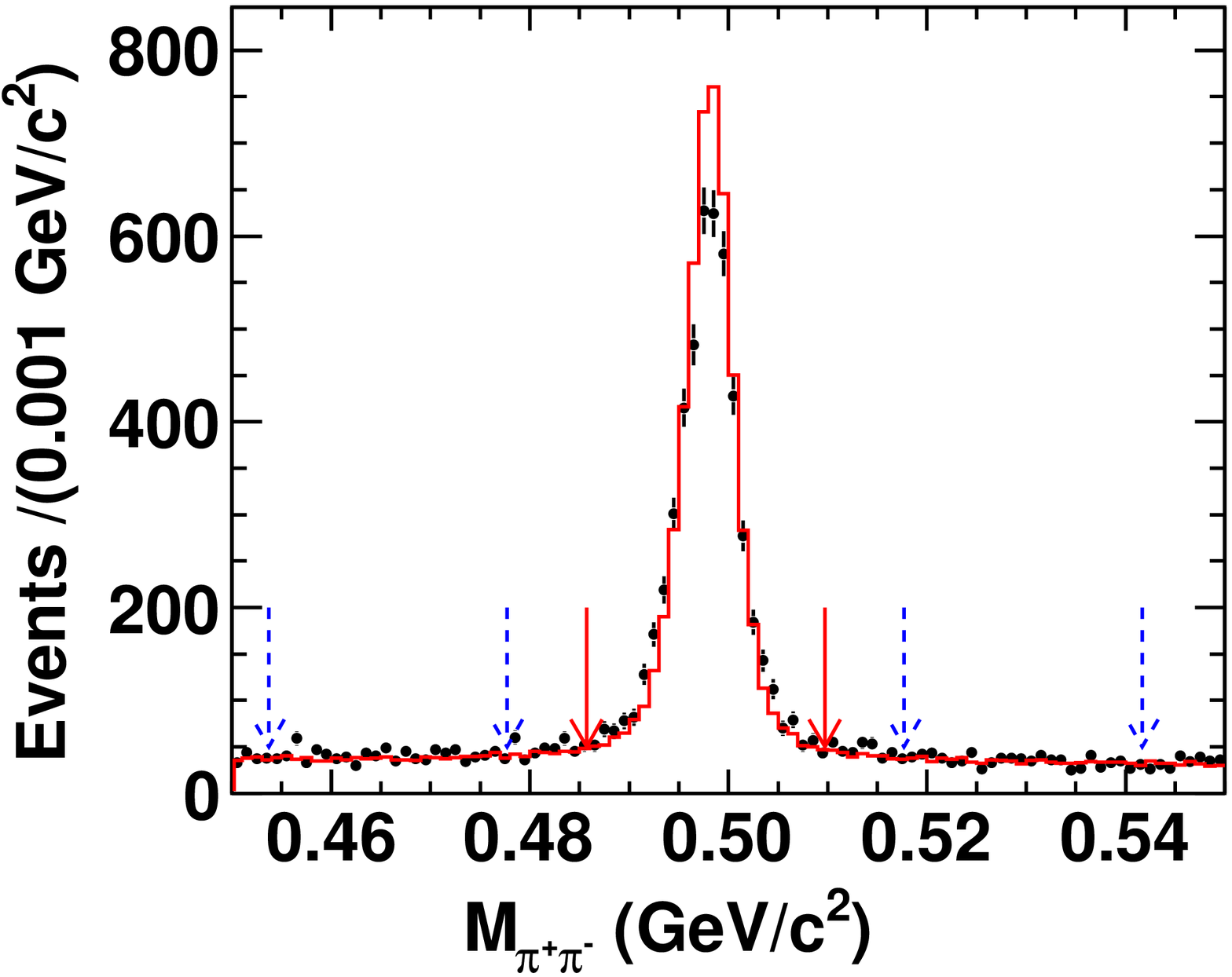}
  \includegraphics[width=2.1in]{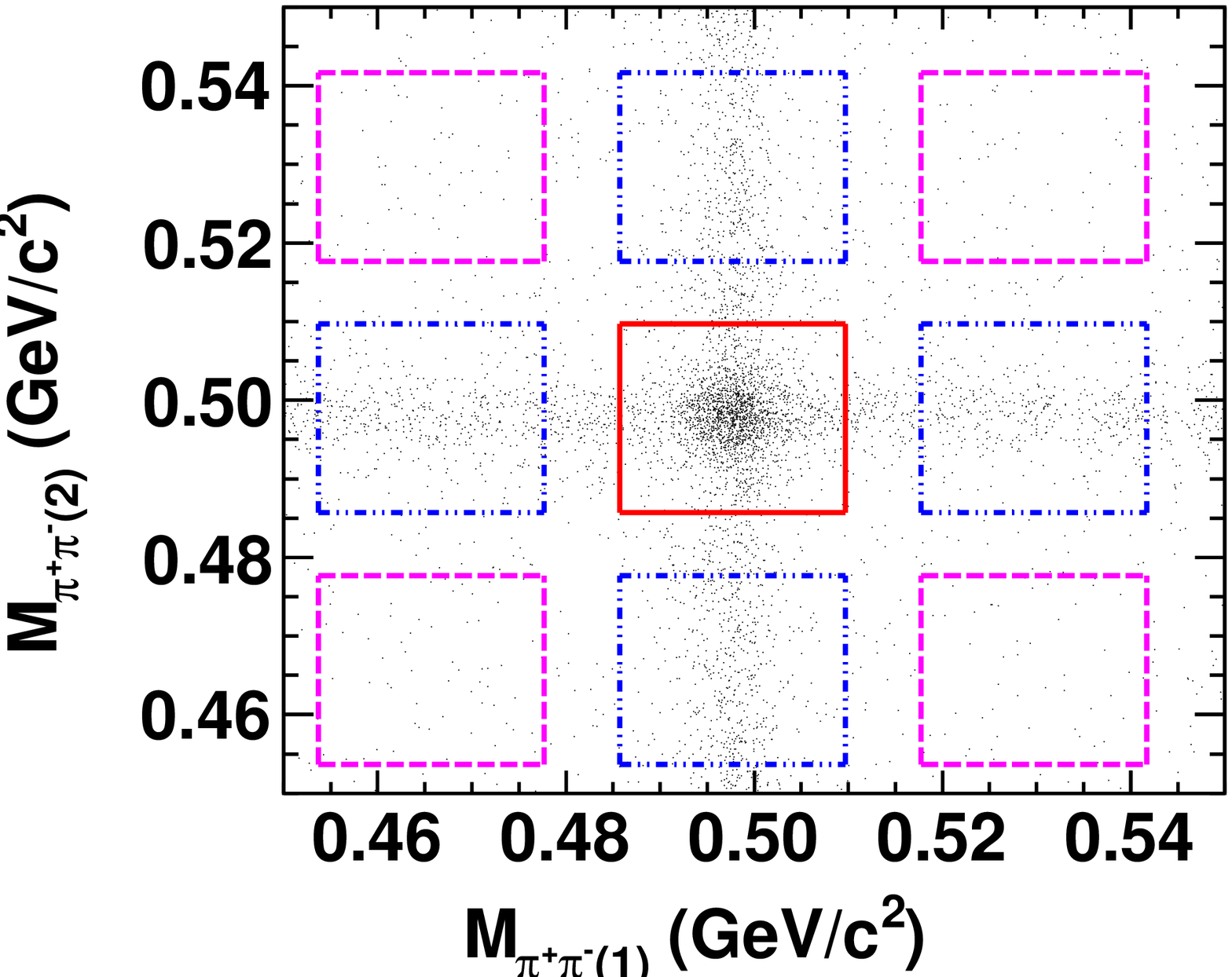}
  \includegraphics[width=2.1in]{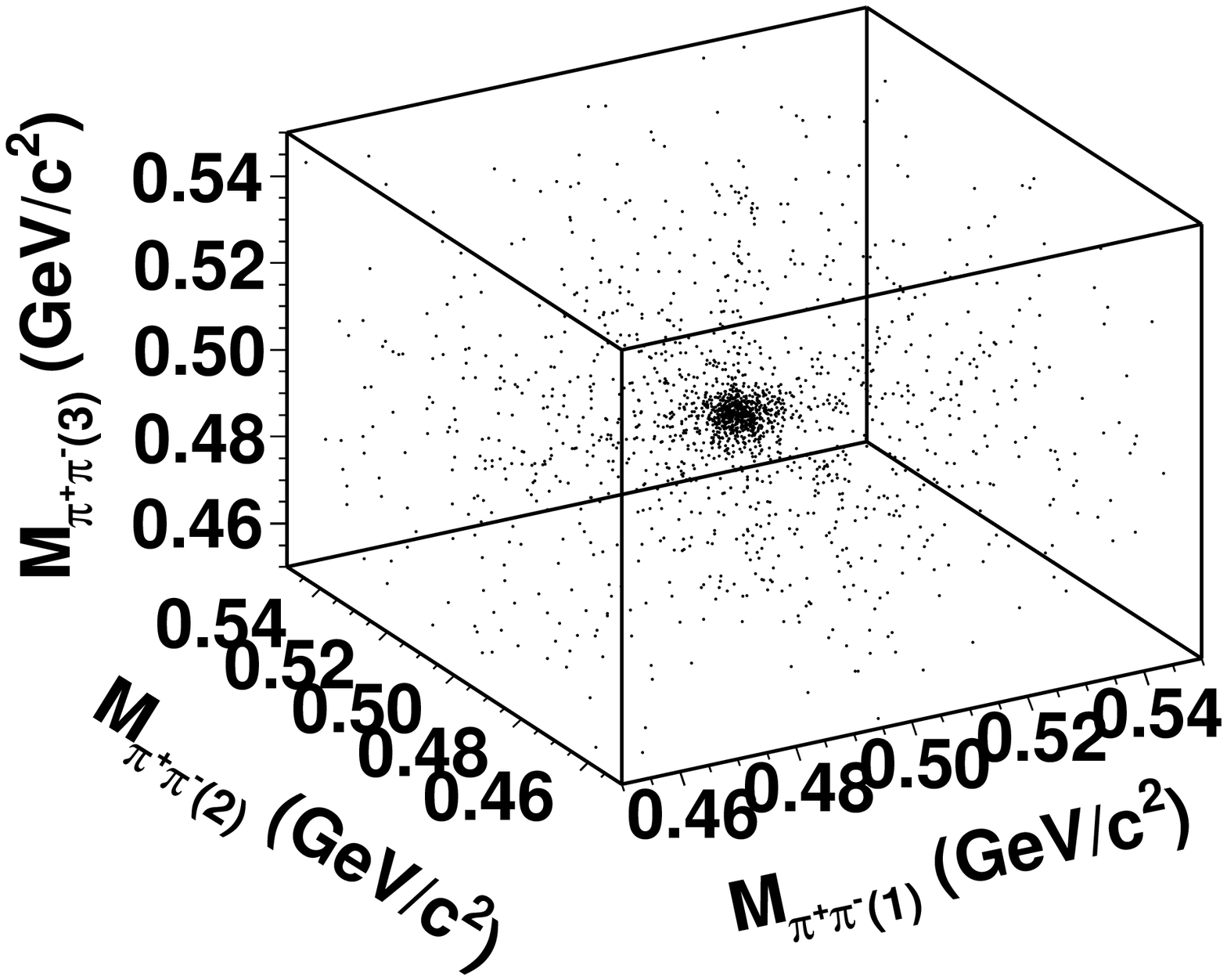}
  \put(-68,116){\bf (c)}
  \put(-223,116){\bf (b)}
  \put(-380,116){\bf (a)}
\caption{
(a)~Comparison of the $M_{\pi^+\pi^-}$ distributions of the $D^0\to K^0_SK^0_S$ candidate events
between data (dots with error bars) and inclusive MC (histogram).
The pairs of the solid~(dashed) arrows denote the $K^0_S$ signal~(sideband) regions.
(b)~Distribution of $M_{\pi^+\pi^-(1)}$ versus $M_{\pi^+\pi^-(2)}$ for the $D^0\to K^0_SK^0_S$ candidate events in data.
(c)~Distribution of $M_{\pi^+\pi^-(1)}$ versus $M_{\pi^+\pi^-(2)}$ versus $M_{\pi^+\pi^-(3)}$ for
the $D^0\to K^0_SK^0_SK^0_S$ candidate events in data.
In these figures,
all selection criteria have been imposed except for the $K^0_S$ mass requirement
and $M_{\rm BC}$ is required to be within 5 MeV/c$^2$ around the $D$ nominal mass~\cite{pdg2014}.
}\label{fig:3mks}
\end{figure*}

The combinatorial $\pi^+\pi^-|_{{\rm non}-K^0_S}$ pairs with invariant mass in $K^0_S$ signal region may also satisfy the $K^0_S$ selection criteria
and contribute peaking background around the $D$ mass in the $M_{\rm BC}$ distribution.
This peaking background is estimated with events in the $K^0_S$ sideband region,
defined as 0.020 $<|M_{\pi^+\pi^-}-M_{K^0_S}|<0.044$ GeV/$c^2$.
Figure~\ref{fig:3mks}(a) shows the comparison of the $M_{\pi^+\pi^-}$ distribution for
$D^0\to K^0_SK^0_S$ candidates in data with the corresponding distribution for the inclusive MC.
In the figure, the solid (dashed) arrows delineate the $K^0_S$ signal (sideband) regions.

In the analyses of the $D^0\to K^0_SK^0_S$, $D^+\to K^0_SK^0_SK^+$ and $K^0_SK^0_S\pi^+$ decays,
two-dimensional (2D) signal and sideband regions are defined.
Figure~\ref{fig:3mks}(b) shows the distribution of $M_{\pi^+\pi^-(1)}$ versus
$M_{\pi^+\pi^-(2)}$ for the $D^0\to K^0_SK^0_S$ candidate events in data.
The solid box, in which both of the $\pi^+\pi^-$ combinations lie in the $K^0_S$ signal regions,
denotes the 2D signal region.
The dot-dashed~(dashed) boxes indicate the 2D sideband 1~(2) regions,
in which one~(two) of the $\pi^+\pi^-$ combinations lie in the $K^0_S$ sideband regions
and the others are in the $K^0_S$ signal region.
For the $D^0\to K^0_SK^0_SK^0_S$ decay, $M_{\pi^+\pi^-(1)}$ versus
$M_{\pi^+\pi^-(2)}$ versus $M_{\pi^+\pi^-(3)}$ of the candidate events in data is shown in Fig.~\ref{fig:3mks} (c).
The region in which all three $\pi^+\pi^-$ combinations lie in the $K^0_S$ signal regions
is taken as the three-dimensional (3D) signal region.
The 3D sideband $i~(i=1,2,3)$ regions denote those in which $i$ of the three $\pi^+\pi^-$
pairs lie in the $K^0_S$ sideband regions
and the rest are located in the $K^0_S$ signal regions.

\begin{table*}[htp]
\centering
\caption{\label{tab:deltaE}$\Delta E$ requirements (in MeV) for data and MC samples.}
\small
\begin{tabular}{lcccc}
  \hline
\multicolumn{1}{c}{Decay modes} & Data & MC\\ \hline
$D^+\to K^0_SK^0_SK^+$&$(-17,+19)$&$(-16,+16)$\\
$D^+\to K^0_SK^0_S\pi^+$&$(-17,+17)$&$(-17,+16)$\\
$D^0\hspace{0.07cm}\to K^0_SK^0_S$&$(-19,+17)$&$(-17,+14)$\\
$D^0\hspace{0.07cm}\to K^0_SK^0_SK^0_S$&$(-14,+16)$&$(-13,+13)$\\ \hline
\end{tabular}
\end{table*}

\begin{figure*}[htp]
  \centering
  \includegraphics[width=5.4in]{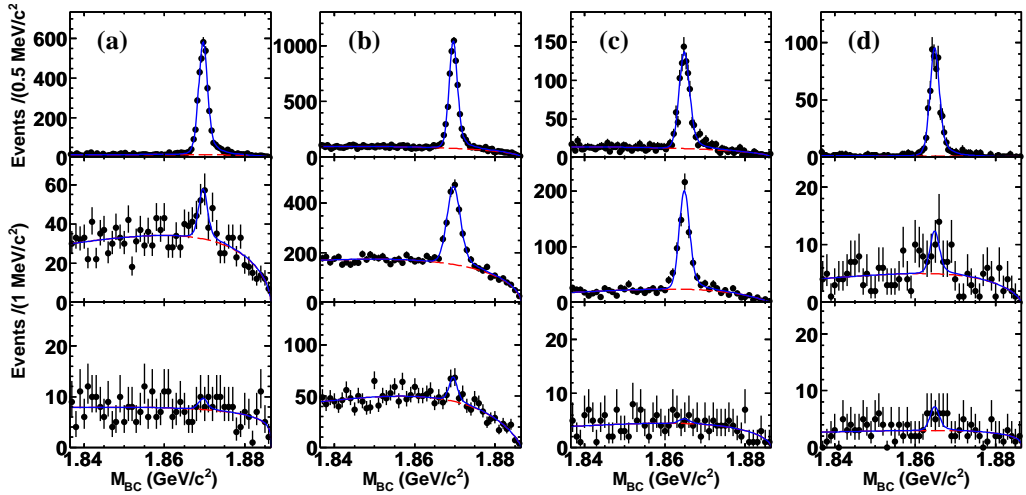}
  \put(-70,170){\bf (d)}
  \put(-163,170){\bf (c)}
  \put(-257,170){\bf (b)}
  \put(-351,170){\bf (a)}
\caption{ Fits to the $M_{\rm BC}$ distributions of the (a) $D^+\to
K^0_SK^0_SK^+$, (b) $D^+\to K^0_SK^0_S\pi^+$, (c) $D^0\to K^0_SK^0_S$ and (d)
$D^0\to K^0_SK^0_SK^0_S$ candidate events.
The dots with error bars are data,
the solid curves are the total fits,
and the dashed curves are the fitted backgrounds.
The first, second and third rows correspond
to the fits to the candidate events in the 2D or 3D signal region,
sideband 1 region and sideband 2 region, respectively.}\label{fig:datafit_Massbc}
\end{figure*}

The resulting $M_{\rm BC}$ distributions of the accepted candidate events
in the 2D or 3D signal region, sideband 1 region and sideband 2 region
are shown in the sub-figures of the first, second and third rows of
Fig.~\ref{fig:datafit_Massbc}, respectively.
By fitting these $M_{\rm BC}$ distributions as shown in Fig.~\ref{fig:datafit_Massbc},
we obtain the fitted yields of $D$ signal
in the 2D or 3D signal region, sideband 1 region and sideband 2 region,
$N_{K^0_S\rm sig}$, $N_{{\rm sb1}}$, $N_{{\rm sb2}}$, which are given in Table~\ref{tab:singletagN_MC}.
In the fits, the $D$ signal is modeled by a MC-simulated shape convoluted with a Gaussian function
with free parameters accounting for the difference of detector resolution between data and MC.  The combinatorial backgrounds are described by an ARGUS function~\cite{ARGUS}
with an endpoint of 1.8865 GeV/$c^2$.
In the $M_{\rm BC}$ fits for the 2D or 3D sideband events, the parameters of the convoluted Gaussian function are fixed at the values determined for the signal region.
For the $D^0\to K^0_SK^0_SK^0_S$ decays, the
peaking backgrounds from sideband 3 region are negligible since few
events survive.

In this analysis, the combinatorial background in the $M_{\pi^+\pi^-}$ distribution are assumed to be flat,
which implies that the ratio of background yields between the $K^0_S$ signal and sideband regions is 0.5.
Thus, the net numbers of the $D^0\to K^0_SK^0_S$, $D^+\to K^0_SK^0_SK^+$ and $K^0_SK^0_S\pi^+$ decays
can be calculated by
\begin{equation}
\label{eq:1}
N_{\rm net} = N_{K^0_S\rm sig} - \frac{1}{2}N_{\rm sb1}+ \frac{1}{4}N_{\rm sb2} - N_{\rm other}^{\rm b},
\end{equation}
and the net number of the $D^0\to K^0_SK^0_SK^0_S$ decays can be calculated by
\begin{equation}
\label{eq:2}
N_{\rm net} = N_{K^0_S\rm sig} - \frac{1}{2}N_{\rm sb1}+\frac{1}{4}N_{\rm sb2}-\frac{1}{8}N_{\rm sb3}-N_{\rm other}^{\rm b},
\end{equation}
where $N_{K^0_S\rm sig}$ and $N_{{\rm sb}i}$
are $D$ signal yields from the fit in the 2D or 3D signal regions and sideband $i$ regions, respectively.
$N_{\rm other}^{\rm b}$ is the normalized number of residual peaking background.
For the $D^+\to K^0_SK^0_SK^+$, $D^+\to K^0_SK^0_S\pi^+$ and $D^0\to K^0_SK^0_SK^0_S$ decays, the residual peaking background is mainly from the events of $D^+\to K^0_SK^0_LK^+$, $D^+\to K^0_SK^0_L\pi^+$ and $D^0\to K^0_SK^0_SK^0_L$ versus $D^-(\bar D^0)\to K^0_SX$ ($X=$ any possible particle combination).
This kind of background peaks around the nominal $D$ mass~\cite{pdg2014}
when the $K^0_S$ from a $D^-(\bar D^0)$ decay has momentum similar to that of a $K^0_L$ produced in 
$D^+(D^0)$ decay.
These peaking backgrounds cannot be modeled by the events from the 2D or 3D sideband region and are estimated by analyzing the inclusive MC sample.
The measured values of $N_{\rm other}^{\rm b}$ and $N_{\rm net}$
are given in Table~\ref{tab:singletagN_MC}.

\section{Branching fractions}

The branching fraction for the hadronic decay $D^{+(0)}\to f$ is determined by
 \begin{equation}\label{equ:branchingfraction}
 {\mathcal B}(D^{+(0)}\to f) = \frac{N_{\rm net}}{2\cdot\sigma_{D^{+}D^{-}~(D^{0}\bar{D}^{0})}\cdot {\mathcal L}\cdot\epsilon},
 \end{equation}
where $N_{\rm net}$ is the net number of $D^{+(0)}\rightarrow f$ decays in data,
$\epsilon$ is the detection efficiency including the branching fraction of $K^0_S \to \pi^+\pi^-$,
$\mathcal L$ is the integrated luminosity of data~\cite{lum}
and $\sigma_{D^{+}D^{-}~(D^{0}\bar{D}^{0})}$
is the $D^{+}D^{-}$ ($D^{0}\bar{D}^{0}$) cross section at the $\psi(3770)$ resonance peak.

The detection efficiencies are determined by analyzing the inclusive MC sample.
In this sample, the signal MC events for $D^+ \to K^0_SK^0_S\pi^+$ are produced
as a mixed sample containing 90\% of the $D^+ \to K^0_SK^{*}(892)^{+},K^{*}(892)^{+}\to K_S^0\pi^+$ decays
and 10\% of the direct three-body decay in phase space $D^+ \to K^0_SK^0_S\pi^+$.
The signal MC events for $D^+ \to K^0_SK^0_SK^+$,
$D^0 \to K^0_SK^0_S$ and $K^0_SK^0_SK^0_S$ are produced using a phase-space model.
Detailed studies show that the momentum and polar-angle distributions of
the daughter particles in data are well modeled by the MC simulation for each decay mode.
By analyzing the inclusive MC sample with the same analysis procedure applied to the data (including the $M_{\rm BC}$ fits and the calculation of the net signal yields),
we obtain the net number of $D$ mesons observed for each decay.
The detection efficiency $\epsilon$ is obtained by dividing the net $D$ signal by the total number 
of signal events, taking into account the efficiency correction discussed in Sect.~\ref{sec:sys}.

Inserting the numbers of $N_{\rm net}$, $\epsilon$, $\mathcal L$,
as well as $\sigma_{D^{+}D^{-}}=(2.882\pm0.018_{\rm stat.}\pm0.042_{\rm sys.})$ nb
or
$\sigma_{D^{0}\bar{D}^{0}}=(3.607\pm0.017_{\rm stat.}\pm0.056_{\rm sys.})$ nb quoted from Ref.~\cite{crsdd-cleo}
into Eq.~(\ref{equ:branchingfraction}), we obtain the branching fraction for each decay,
as listed in Table~\ref{tab:singletagN_MC}, where the uncertainties are statistical only.

\begin{table*}[htp]
\centering
\caption{\label{tab:singletagN_MC}
Input quantities and results for the determination of the branching fractions as described in the text.
The uncertainties are statistical only.
}
\small
\begin{tabular}{lcccccccc} \hline
\multicolumn{1}{c}{Decay modes} & $N_{K^0_S\rm sig}$&  $N_{\rm sb1}$&  $N_{\rm sb2}$& $N_{\rm sb3}$ &$N_{\rm other}^{\rm b}$ &$N_{\rm net}$&  $\epsilon$ (\%) &$\mathcal B$ ($\times 10^{-4}$)\\ \hline
$D^+\to K^0_SK^0_SK^+$&$3616\pm66$&$\hspace{0.33cm}97\pm19$&$\hspace{0.15cm}6\pm\hspace{0.15cm}8$&--&$18\pm2$&$3551\pm67$&$\hspace{0.15cm}8.27\pm0.04$&$25.4\pm0.5$\\
$D^+\to K^0_SK^0_S\pi^+$&$5643\pm88$&$1464\pm68$&$69\pm19$&--&$31\pm3$&$4897\pm94$&$10.72\pm0.04$&$27.0\pm0.5$\\
$D^0\hspace{0.07cm}\to K^0_SK^0_S$&$\hspace{0.15cm}888\pm36$&$\hspace{0.15cm}626\pm31$&$\hspace{0.15cm}3\pm\hspace{0.15cm}6$&--&\hspace{0.15cm}0&
$\hspace{0.15cm}576\pm39$&$16.28\pm0.30$&$\hspace{0.15cm}1.67\pm0.11$\\
$D^0\hspace{0.07cm}\to K^0_SK^0_SK^0_S$&$\hspace{0.15cm}622\pm27$&$\hspace{0.30cm}24\pm\hspace{0.15cm}8$&$14\pm\hspace{0.15cm}6$&0
&$\hspace{0.15cm}16\pm3$&$\hspace{0.15cm}597\pm27$&$\hspace{0.15cm}3.92\pm0.05$&$\hspace{0.15cm}7.21\pm0.33$\\ \hline
\end{tabular}
\end{table*}

\section{Systematic uncertainty}
\label{sec:sys}

Table~\ref{tab:relsysuncertainties} shows the systematic uncertainties
in the branching fraction measurements.
Each of them, estimated relative to the measured branching
fraction, is discussed below.

\begin{itemize}
\item
{\it MC statistics}:
The uncertainties due to the limited MC statistics are 0.5\%, 0.4\%, 1.8\% and 1.3\%
for $D^+\to K^0_SK^0_SK^+$, $D^+\to K^0_SK^0_S\pi^+$, $D^0\to K^0_SK^0_S$ and $D^0\to K^0_SK^0_SK^0_S$, respectively.

\item
{\it Luminosity of data}:
The uncertainty in the quoted integrated luminosity of data is 0.5\%~\cite{lum}.

\item
{\it $D\bar D$ cross section}:
The uncertainties of the quoted $D^+D^-$ and $D^0\bar D^0$
cross sections are 1.6\%~\cite{crsdd-cleo}.

\item
{\it ${\mathcal B}(K^0_S\to \pi^+\pi^-)$}:
The uncertainty of the quoted branching fraction for $K^0_S\to \pi^+\pi^-$ is 0.1\%~\cite{pdg2014}.

\item
{\it $K_S^0$ reconstruction}:
The $K_{S}^{0}$ reconstruction efficiency
has been studied as a function of momentum by using the control samples
$J/\psi\to K^{*}(892)^{\mp}K^{\pm}$ and $J/\psi\to \phi K_S^{0}K^{\pm}\pi^{\mp}$.
Small data-MC efficiency differences are found and presented in Ref.~\cite{sysks}.
To correct the $K^0_S$ reconstruction efficiency, a piecewise fit
to these differences as a function of $K^0_S$ momentum is performed.
For the efficiencies of detecting the decays $D^+\to K^0_SK^0_SK^+$,
$D^+\to K^0_SK^0_S\pi^+$, $D^0\to K^0_SK^0_S$ and $D^0\to K^0_SK^0_SK^0_S$,
the momentum weighted differences associated with $K^0_S$ reconstruction between
data and MC are determined to be
$(+3.9\pm1.9)\%$,
$(+3.0\pm1.4)\%$,
$(+1.8\pm0.8)\%$ and
$(+5.9\pm2.8)\%$, respectively,
where the uncertainties are statistical.
These corrections are applied to the detection efficiencies, after which
only the statistical uncertainties of the differences are retained.
On average, the residual uncertainty for each $K^0_S$ is no more than 1.0\%.
Furthermore, the difference of the momentum-weighted efficiencies
between data and MC from the different fits, which is 1.0\% per
$K^0_S$, is included as an additional uncertainty. Finally, we
assign 1.5\% per $K^0_S$ as the systematic uncertainty for the 
reconstruction efficiency .

\item
{{\it Tracking} [PID] for $K^+(\pi^+)$}:
The tracking [PID] efficiencies for $K^+$ and $\pi^+$
are investigated using doubly tagged $D\bar D$ hadronic events.
The difference of momentum weighted efficiencies between data and MC of the tracking [PID]
are determined to be
$(+2.1\pm0.4)\%$ [$(-0.3\pm0.1)\%$] for the $K^+$ in the $D^+\to K^0_SK^0_SK^+$ decay
and
$(+0.4\pm0.3)\%$ [$(-0.3\pm0.1)\%$] for the $\pi^+$ in the $D^+\to K^0_SK^0_S\pi^+$ decay,
where the uncertainties are statistical.
After correcting the detection efficiencies by these differences,
we take 0.5\% [0.5\%] as the systematic uncertainties in
tracking [PID] for the $K^+$ and $\pi^+$, respectively.

\item
{\it $M_{\rm BC}$ fit}:
In order to estimate the systematic uncertainty associated with the $M_{\rm BC}$ fit, we repeat
the measurements by varying the fit range ($(1.8415,1.8865)$ GeV/$c^2$),
signal shape (with different MC matching requirements)
and endpoint of the ARGUS function ($\pm0.2$ MeV/$c^2$).
Quadratically summing the changes of the branching fractions
yields 2.1\%, 1.0\%, 4.2\% and 2.7\% for
$D^+\to K^0_SK^0_SK^+$, $D^+\to K^0_SK^0_S\pi^+$, $D^0\to K^0_SK^0_S$
and $D^0\to K^0_SK^0_SK^0_S$,
which are assigned as the relevant systematic uncertainties.

\item
{\it $\Delta E$ requirement}:
To investigate the systematic uncertainty associated with the $\Delta E$ requirement,
we repeat the measurements using alternative $\Delta E$
requirements of $\pm (4,5,6)$ times the resolution around the $\Delta E$ peaks.
The maximum changes of the branching fractions, 2.0\%, 1.5\%, 2.0\% and 1.5\% for
$D^+\to K^0_SK^0_SK^+$, $D^+\to K^0_SK^0_S\pi^+$, $D^0\to K^0_SK^0_S$
and $D^0\to K^0_SK^0_SK^0_S$,
are taken as the associated systematic uncertainties.

\item
{\it Normalization of peaking backgrounds}:
In the nominal analysis, the normalization factor for the peaking backgrounds,
which is the ratio of background yields between the $K^0_S$ signal and sideband regions,
has been assumed to be 0.5.
The branching fractions are recalculated with alternative normalization factors 
determined by MC simulation. The corresponding changes on the branching fractions,
0.5\%, 1.4\%, 2.4\% and 0.7\% for
$D^+\to K^0_SK^0_SK^+$, $D^+\to K^0_SK^0_S\pi^+$, $D^0\to K^0_SK^0_S$ and $D^0\to K^0_SK^0_SK^0_S$,
are assigned as the systematic uncertainties associated with the peaking background (PBKG) normalization.
On the other hand, the uncertainties of the residual peaking backgrounds are dominated by the uncertainties of the 
input branching fractions for $D^-(\bar D^0)\to K^0_SX$,
which contribute additional uncertainties of 0.1\%, 0.1\% and 0.4\% for the measured branching fractions for $D^+\to K^0_SK^0_SK^+$, $D^+\to K^0_SK^0_S\pi^+$ and $D^0\to K^0_SK^0_SK^0_S$, respectively.

\item
{\it $K^0_S$ sideband}:
To evaluate the systematic uncertainty due to the choice of $K^0_S$ sideband region,
we remeasure the branching fractions after shifting the $K^0_S$ sideband by $\pm 2$ MeV/$c^2$. 
The corresponding maximum changes in the branching fraction, which are 0.5\%, 0.5\%, 2.0\% and 1.0\%
for $D^+\to K^0_SK^0_SK^+$, $D^+\to K^0_SK^0_S\pi^+$, $D^0\to K^0_SK^0_S$ and $D^0\to K^0_SK^0_SK^0_S$, respectively,
are taken as the systematic uncertainties.

\item
{\it MC modeling}:
For the three-body decays,
we examine the reweighted detection efficiencies by including
the possible sub-resonances $a_0(980)$ and $f_0(980)$ in the signal MC samples.
The maximum change of the reweighted detection efficiencies,
1.0\%, is taken as the systematic uncertainty in MC modeling.
\end{itemize}

Adding all of above systematic uncertainties in quadrature, we obtain the total
systematic uncertainties of 4.7\%, 4.4\%, 6.8\% and 6.1\% for
$D^+\to K^0_SK^0_SK^+$, $D^+\to K^0_SK^0_S\pi^+$, $D^0\to K^0_SK^0_S$ and $D^0\to K^0_SK^0_SK^0_S$, respectively.

\begin{table*}[htp]
\centering
\caption{\label{tab:relsysuncertainties}Systematic uncertainties (\%) in the branching fraction measurements.}
\begin{small}
\begin{tabular}{ccccc}
  \hline
  Sources &$D^+\to K^0_SK^0_SK^+$&$D^+\to K^0_SK^0_S\pi^+$&$D^0\to K^0_SK^0_S$&$D^0\to K^0_SK^0_SK^0_S$\\
  \hline
MC statistics              & 0.5 & 0.4 & 1.8 & 1.3 \\
Luminosity of data         & 0.5 & 0.5 & 0.5 & 0.5 \\
$D\bar D$ cross section    & 1.6 & 1.6 & 1.6 & 1.6 \\
${\mathcal B}(K^0_S\to \pi^+\pi^-)$ & 0.2 & 0.2 & 0.2 & 0.3 \\
$K_S^0$ reconstruction     & 3.0 & 3.0 & 3.0 & 4.5 \\
Tracking for $K^+(\pi^+)$  & 0.5 & 0.5 & --  & --  \\
PID for $K^+(\pi^+)$       & 0.5 & 0.5 & --  & --  \\
$M_{\rm BC}$ fit           & 2.1 & 1.0 & 4.2 & 2.7 \\
$\Delta E$ requirement     & 2.0 & 1.5 & 2.0 & 1.5 \\
PBKG normalization         & 0.5 & 1.4 & 2.4 & 0.8 \\
$K_S^0$ sideband           & 0.5 & 0.5 & 2.0 & 1.0 \\
MC modeling                  & 1.0 & 1.0 & -- & 1.0 \\
\hline
Total            &4.7 & 4.4 & 6.8 & 6.1 \\ \hline
\end{tabular}
\end{small}
\end{table*}

\section{Summary}

In summary, by analyzing 2.93 $\rm fb^{-1}$ of data collected at $\sqrt s=$ 3.773 GeV
with the BESIII detector, we measure the branching fractions for the hadronic decays
$D^+\to K^0_SK^0_SK^+$, $D^+\to K^0_SK^0_S\pi^+$, $D^0\to K^0_SK^0_S$ and
$D^0\to K^0_SK^0_SK^0_S$ using a single-tag method.
Table~\ref{tab:combranch} presents the comparisons of the measured branching fractions
with the PDG values~\cite{pdg2014}.
The branching fraction for $D^+\to K^0_SK^0_S\pi^+$ is measured for the first time
and the others are consistent with previous
measurements, but with much improved precision.
We also determine the branching fraction ratios
${\mathcal B}(D^+\to K^0_SK^0_SK^+)/{\mathcal B}(D^+\to K^0_SK^0_S\pi^+)=0.941\pm0.025_{\rm stat.}\pm0.040_{\rm sys.}$
and
${\mathcal B}(D^0\to K^0_SK^0_S)/{\mathcal B}(D^0\to K^0_SK^0_SK^0_S)=0.232\pm0.019_{\rm stat.}\pm0.016_{\rm sys.},$
in which the systematic uncertainties in the $D^+D^-$ (or $D^0\bar D^0$) cross section,
the integrated luminosity of data, as well as the reconstruction
efficiencies and the branching fractions of the two $K^0_S$ mesons cancel.
The results in this analysis provide helpful experimental data to probe for the interplay between the weak and strong interactions in charmed meson decay~\cite{plb193_331,prd45_4113,prd60_014014,prd64_034010}.
In addition, the measured branching fraction for the two-body decay $D^0\to K^0_SK^0_S$ can also help to understand SU(3)-flavor symmetry breaking effects in $D$ meson decays~\cite{Kwong,prd86_036012,Grossman,prd92_014004,prd92_014032}.
\begin{table*}[htp]
\centering
\caption{\label{tab:combranch}
Comparisons of the branching fractions (in $10^{-4}$) measured in this work with the PDG values~\cite{pdg2014}.}
\small
\begin{tabular}{lcc}  \hline
\multicolumn{1}{c}{Decay modes}        & This work        &   PDG               \\
  \hline
$D^+\to K^0_SK^0_SK^+$&  $ 25.4 \pm 0.5\hspace{0.15cm}\pm 1.2\hspace{0.03cm}~$  &  $45\pm20$               \\
$D^+\to K^0_SK^0_S\pi^+$&  $ 27.0 \pm 0.5\hspace{0.15cm}\pm 1.2\hspace{0.03cm}~$  &  --     \\
$D^0\hspace{0.07cm}\to K^0_SK^0_S$&  $ 1.67 \pm 0.11 \pm 0.11       $  &  $1.7\pm0.4$               \\
$D^0\hspace{0.07cm}\to K^0_SK^0_SK^0_S$&  $ 7.21 \pm 0.33 \pm 0.44 $  &  $9.1\pm 1.3 $               \\
  \hline
\end{tabular}
\end{table*}

\section{Acknowledgements}
The BESIII collaboration thanks the staff of BEPCII and the IHEP computing center
for their strong support. This work is supported in part by National Key Basic
Research Program of China under Contract Nos. 2009CB825204 and 2015CB856700; National Natural
Science Foundation of China (NSFC) under Contracts Nos. 10935007, 11235011, 11305180, 11322544,
11335008, 11425524, 11475123; the Chinese Academy of Sciences (CAS) Large-Scale Scientific
Facility Program; the CAS Center for Excellence in Particle Physics (CCEPP);
the Collaborative Innovation Center for Particles and Interactions (CICPI);
Joint Large-Scale Scientific Facility Funds of the NSFC and CAS under Contracts
Nos. U1232201, U1332201, U1532101, U1532257, U1532258; CAS under Contracts Nos. KJCX2-YW-N29, KJCX2-YW-N45;
100 Talents Program of CAS; National 1000 Talents Program of China; INPAC and
Shanghai Key Laboratory for Particle Physics and Cosmology; German Research Foundation DFG under Contracts Nos. Collaborative Research Center CRC 1044, FOR 2359;
Istituto Nazionale di Fisica Nucleare, Italy;
Koninklijke Nederlandse
Akademie van Wetenschappen (KNAW) under Contract No. 530-4CDP03; Ministry of
Development of Turkey under Contract No. DPT2006K-120470; The Swedish Research
Council; U. S. Department of Energy under Contracts Nos. DE-FG02-05ER41374,
DE-SC-0010504, DE-SC0012069, DESC0010118; U.S. National Science Foundation;
University of Groningen (RuG) and the Helmholtzzentrum fuer Schwerionenforschung
GmbH (GSI), Darmstadt; WCU Program of National Research Foundation of Korea under
Contract No. R32-2008-000-10155-0.

\end{multicols}

\begin{thebibliography}{**}

\bibitem{pdg2014}
C. Patrignani {\it et al.} (Particle Data Group), Chin. Phys. C {\bf 40}, 100001 (2016).

\bibitem{plb193_331}
X. Y. Pham {\it et al.}, Phys. Lett. B {\bf 193}, 331 (1987).

\bibitem{prd45_4113}
R. E. Karlsen and M. D. Scadron {\it et al.}, Phys. Rev. D {\bf 45}, 4113 (1992).

\bibitem{prd60_014014}
Y. S. Dai {\it et al.}, Phys. Rev. D {\bf 60}, 014014 (1999).

\bibitem{prd64_034010}
J. O. Eeg {\it et al.}, Phys. Rev. D {\bf 64}, 034010 (2001).

\bibitem{Kwong}
W. Kwong and S. P. Rosen, Phys. Lett. B 298, 413 (1993).

\bibitem{prd86_036012}
H. N. Li {\it et al.}, Phys. Rev. D {\bf 86}, 036012 (2012).

\bibitem{Grossman}
Y. Grossman and D. J. Robinson, JHEP 1304, 67 (2013).

\bibitem{prd92_014004}
S. M\"{u}ller {\it et al.}, Phys. Rev. D {\bf 92}, 014004 (2015).

\bibitem{prd92_014032}
A. Biswas {\it et al.}, Phys. Rev. D {\bf 92}, 014032 (2015).

\bibitem{lum}M. Ablikim \emph{et al.} (BESIII Collaboration), Chin. Phys. C {\bf 37}, 123001 (2013);
Phys. Lett. B {\bf 753}, 629 (2016).

\bibitem{bes3}
M. Ablikim {\it et al}. (BESIII Collaboration), Nucl. Instrum. Meth. A {\bf 614}, 345 (2010).

\bibitem{zpc46_9}
H. Albrecht {\it et al.} (ARGUS Collaboration), Z. Phys. C {\bf 46}, 9 (1990).

\bibitem{zpc64_375}
H. Albrecht {\it et al.} (ARGUS Collaboration), Z. Phys. C {\bf 64}, 375 (1994).

\bibitem{prd44_3383}
R. Ammar {\it et al.} (CLEO Collaboration), Phys. Rev. D {\bf 44}, 3383 (1991).

\bibitem{prd44_4211}
D. M. Asner {\it et al.} (CLEO Collaboration), Phys. Rev. D {\bf 54}, 4211 (1996).

\bibitem{plb607_59}
J. M. Link {\it et al.} (FOCUS Collaboration), Phys. Lett. B {\bf 607}, 59 (2005).

\bibitem{crsdd-cleo}
G. Bonvicini {\it et al.} (CLEO Collaboration), Phys. Rev. D {\bf 89}, 072002 (2014).

\bibitem{geant4}
S. Agostinelli {\it et al.} (GEANT4 Collaboration),
Nucl. Instrum. Meth. A {\bf 506}, 250 (2003).

\bibitem{kkmc}
S. Jadach, B. F. L. Ward and Z. Was,
Comp. Phys. Commu. {\bf 130}, 260 (2000);
Phys. Rev. D {\bf 63}, 113009 (2001).

\bibitem{isr}
E. A. Kureav and V. S. Fadin, Sov. J. Nucl. Phys. {\bf 41}, 466 (1985),
Yad. Fiz. {\bf 41}, 733 (1985).

\bibitem{photons}
E. Richter-Was, Phys. Lett. B {\bf 303}, 63 (1993).

\bibitem{evtgen}
D. J. Lange, Nucl. Instrum. Meth. A {\bf 462}, 152 (2001);
R. G. Ping, Chin. Phys. C {\bf 32}, 599 (2008).

\bibitem{lundcharm}
J. C. Chen {\it et al.}, Phys. Rev. D {\bf 62}, 034003
(2000).

\bibitem{ARGUS} H. Albrecht {\it et al.} (ARGUS Collaboration), Phys. Lett. B {\bf 241}, 278 (1990).

\bibitem{sysks}
M. Ablikim {\it et al}. (BESIII Collaboration),
Phys. Rev. D {\bf 92}, 112008 (2015).

\end{thebibliography}
\end{document}